\documentclass{article}
\usepackage{lscape}
\usepackage{amssymb}

\usepackage{epsfig}

\usepackage{graphicx,psfrag,amsmath}

\begin{document}

\def\beq#1\eeq{\begin{equation}#1\end{equation}}
\def\beql#1#2\eeql{\begin{equation}\label{#1}#2\end{equation}}

\def\bea#1\eea{\begin{eqnarray}#1\end{eqnarray}}
\def\beal#1#2\eeal{\begin{eqnarray}\label{#1}#2\end{eqnarray}}

\newcommand{\Z}{{\mathbb Z}}
\newcommand{\N}{{\mathbb N}}
\newcommand{\C}{{\mathbb C}}
\newcommand{\Cs}{{\mathbb C}^{*}}
\newcommand{\R}{{\mathbb R}}
\newcommand{\intT}{\int_{[-\pi,\pi]^2}dt_1dt_2}
\newcommand{\cC}{{\mathcal C}}
\newcommand{\cI}{{\mathcal I}}
\newcommand{\cN}{{\mathcal N}}
\newcommand{\cE}{{\mathcal E}}
\newcommand{\cA}{{\mathcal A}}
\newcommand{\xdT}{\dot{{\bf x}}^T}
\newcommand{\bDe}{{\bf \Delta}}

\def\ket#1{\left| #1\right\rangle }
\def\bra#1{\left\langle #1\right| }
\def\braket#1#2{\left\langle #1\vphantom{#2}
  \right. \kern-2.5pt\left| #2\vphantom{#1}\right\rangle }
\newcommand{\gme}[3]{\bra{#1}#3\ket{#2}}
\newcommand{\ome}[2]{\gme{#1}{#2}{\mathcal{O}}}
\newcommand{\spr}[2]{\braket{#1}{#2}}
\newcommand{\eq}[1]{Eq\,\ref{#1}}
\newcommand{\xp}[1]{e^{#1}}

\def\limfunc#1{\mathop{\rm #1}}
\def\Tr{\limfunc{Tr}}

\def\dr{detector }
\def\drn{detector}
\def\dtn{detection }
\def\dtnn{detection}

\def\pho{photon }
\def\phon{photon}
\def\phos{photons }
\def\phosn{photons}
\def\mmt{measurement }
\def\an{amplitude}
\def\a{amplitude }
\def\co{coherence }
\def\con{coherence}

\def\st{state }
\def\stn{state}
\def\sts{states }
\def\stsn{states}

\def\cow{"collapse of the wavefunction"}
\def\de{decoherence }
\def\den{decoherence}
\def\dm{density matrix }
\def\dmn{density matrix}

\newcommand{\mop}{\cal O }
\newcommand{\dt}{{d\over dt}}
\def\qm{quantum mechanics }
\def\qms{quantum mechanics }
\def\qml{quantum mechanical }

\def\qmn{quantum mechanics}
\def\mmtn{measurement}
\def\pow{preparation of the wavefunction }

\def\me{ L.~Stodolsky }
\def\T{temperature }
\def\Tn{temperature}
\def\t{time }
\def\tn{time}
\def\wfs{wavefunctions }
\def\wf{wavefunction }
\def\wfn{wavefunction} 
\def\wfsn{wavefunctions}
\def\wvp{wavepacket }
\def\pa{probability amplitude } 
\def\sy{system } 
\def\sys{systems }
\def\syn{system} 
\def\sysn{systems} 
\def\ha{hamiltonian }
\def\han{hamiltonian}
\def\rh{$\rho$ }
\def\rhn{$\rho$}
\def\op{$\cal O$ }
\def\opn{$\cal O$}
\def\yy{energy }
\def\yyn{energy}
\def\yys{energies }
\def\yysn{energies}
\def\pz{$\bf P$ }
\def\pzn{$\bf P$}
\def\pl{particle }
\def\pls{particles }
\def\pln{particle}
\def\plsn{particles}

\def\plz{polarization  }
\def\plzs{polarizations }
\def\plzn{polarization}
\def\plzsn{polarizations}

\def\sctg{scattering }
\def\sctgn{scattering}

\def\prob{probability }
\def\probn{probability}

\def\om{\omega} 

\def\hf{\tfrac{1}{2}}

\def\zz{neutrino }
\def\zzn{neutrino}
\def\zzs{neutrinos }
\def\zzsn{neutrinos}

\def\zn{neutron }
\def\znn{neutron}
\def\zns{neutrons }
\def\znsn{neutrons}

\def\csss{cross section }
\def\csssn{cross section}

\def\jp{$J/\psi$ }
\def\b{$B^o$\,}
\def\k{$K^o$\,}
\def\ks{$K_s$\,}
\def\kl{$K_l$\,}
\def\bb{$\bar B^o$\,}
\def\kb{$\bar K^{o^{~}}$\,}

\def\jpn{J/\psi}
\def\bn{B^o}
\def\kn{K^o}
\def\ksn{K_s}
\def\kln{K_l}
\def\bbn{\bar {B^o }}
\def\kbn{\bar {K^o }}

\def\sm{Standard Model }
\def\smn{Standard Model}

\def\sgx{\sigma_1}
\def\sgy{\sigma_2}
\def\sgz{\sigma_3}

\def\fth{\tfrac{1}{4}}

\title{    Observability of
`Cascade Mixing'  in\\ $\bn \to$\jp\k  }

\author{
 L. Stodolsky\\
Max-Planck-Institut f\"ur Physik
(Werner-Heisenberg-Institut)\\
F\"ohringer Ring 6, 80805 M\"unchen, Germany}

\maketitle

\begin{abstract}
 In high statistics observations of $B^o\to J/\psi\,
K^o$ originating from the process $\Upsilon(4S)\to$\b\bb\, it
should be
possible to observe
`cascade mixing',  where one mixing \pln, the \b, turns into
another, the \k.
This is possible despite the difficulty that the length of the beam
crossing
region makes a precise
 definition of the
primary vertex impossible. This difficulty is circumvented by using
 an `away side' tag to specify the initial time. We
review the formalism for describing such processes,
 and first apply it to simple \b mixing,
 noting it 
gives a transparent description for CP and T asymmetries. In
particular we  show that three
different asymmetries of the CP and T type, with neglect of direct
CP violation, are given by the same expression.

For ``cascade mixing"
 we present predictions for processes
of the type $B_i \to K_j$ via \jp, where in the limit 
of no direct CP violation each state i or j is determined by a
simple tag. There are 16  such simple measureable processes,
involving  10 functions of the two time intervals involved. The
coefficients of the functions are different for each of the
processes and are given in terms of the mass splitting and the 
CP, T violating parameter of the \b mass matrix $m_2$.
The results presented here are just consequences of the \qm of \pl
mixing and do
not involve any particular model of CP violation.

\end{abstract}

\section{Introduction}\label{intro}
 Some years ago the idea of ``double'' or ``cascade'' mixing,
where one mixing \sy turns into another via  a decay process was
introduced \cite{az},  \cite{us}, \cite{branco}. Such
processes would exhibit amusing \qml interference effects and
could also provide information on certain properties of the
interfering
\sys.

 The most discussed process of this type 
involves decays of $B^o$ mesons to a $J/\psi\, K^o$ state. The
first
mixing \sy would be the  $B^o$ and the second the $K^o$, while the
decay through the $J/\psi$ provides a kind of ``regeneration" or 
``filter" for the initial state of the \k. This can provide a new
and interesting tool in the manipulation of mixing \sysn, analogous
to the passing of a \k beam through a material of variable
thickness, but where
additionally the quantum numbers of the decay, such as for the p-
wave \jp,  play a role in determining the evolution. Thus in
addition to \jp, one may consider processes mediated by other
\plsn,
with different predictions for the
behavior of the \k oscillations.

Since the original proposals considerable time has elapsed and much
data has been accumulated and is promised for $B^o$ processes. It
would therefore
appear appropriate to reconsider the possibility of studying
``cascade mixing" experimentally. 

 However, there would
appear to be a difficulty. Study of the process involves the
determination of 
 two time differences: $\tau(1,0)$, the proper time for the
interval
between the \b creation and the decay to \jp \k ; and $\tau(2,1)$,
the proper time for the interval between this decay  and
the final \k decay.
The difficulty is that primary vertex of the process,
seemingly necessary to determine $\tau(1,0)$, 
is not well determined in space by the apparatus. For example at
Belle, where one studies $e^+e^-\to \Upsilon(4S)\to$ \b\bb ,  the
length
of the beam
crossing region, is
on the order of several  cm. Such distances are much larger than 
$(\Delta m_B)^{-1}\sim 0.5\times 10^{-12} s \sim 1.5\times
10^{-2}cm $ \cite{booklet} relevant for \b
oscillations, making it
seem unrealistic to observe
any oscillatory effects in  $\tau(1,0)$; and it might be feared
that the
rapid oscillations in  $\tau(1,0)$  will wash out
any interference effects at all.

\subsection{`Away side' tag}

Nevertheless,  in  the $\Upsilon(4S) \to$ \b\b process  there is a
way around this
difficulty. Namely, one may use the method of the ``away side tag''
to determine
the initial time. 
In this method, one uses \cite{lipkin} the p-wave nature of the 
$\Upsilon(4S) \to$ \b\b decay to say that if one 
\b is
observed to be in a given state, then the other member of the pair
must be in the
opposite, orthogonal state. For example, if one meson is observed
to be in the \b state then the other one --at the same time-- is in
the \bb\, state.  \footnote{Although `at the same time' sounds like
a frame-dependent, non-covariant specification, we have explained
elsewhere \cite{coll} that the procedure can be put in covariant
language and that the $\Upsilon$ rest frame is indeed the correct
frame for the procedure. }  For our present purposes  `at the same
time' is the important point. This implies that a measurement of
the `away side' specifies not only the state of the meson
under consideration, but also the {\it time} when it was in this
state--{\it without having to know the original
vertex}. The possibility of such measurements  to sufficient
accuracy has been demonstrated by the Belle \cite{belle} and Babar
\cite{babar} groups in their studies of CP violation.

In the following we shall refer to the \b opposite to the
initial `away side' tag    and its subsequent development, as
the `same side', since this is the \sy we wish to study.

\section{Formalism }

 We use the formalism
introduced in ref\,\cite{us}, which we briefly review here.
One operates in a two-dimensional vector space spanned by \b and
\bb before the \jp decay  or by \k and \kb after the decay. All
quantities are either `spinors' in this space, or 2 x 2
matrices. Examples of `spinors'  are states like  $B_1$  or $K_s$ 
mesons.  The propagation in time of these states, or  the
conversion
of a \b to a \k state  via the decay, are given by matrices. We
shall also use density matrices $\rho$ to describe the initial or
final states of the two-state \syn.

Either for the ordinary one-time mixing,  or for the  "cascade
mixing"  with two times,
 the expressions needed will have the form

\beql{form}
 Tr[\rho(b){\cal M}\rho(a){\cal M}^\dagger]\,.
\eeql
This expression represents the probability amplitude squared to
begin with a state $a$ of the two-state \sy and to end with 
a state $b$ of the two-state \syn.  For ordinary  \b mixing $a$ and
$b$ are different (or perhaps the same ) \b states and $\cal M$
 will depend only on one time difference. For `cascade mixing' $a$
is a \b state and $b$ is a \k state, and the expression
  depends on the two time intervals. To convert \eq{form} into an
experimental rate for a given final channel $\alpha$, it is
necessary to multiply it by a rate constant  $\Gamma(b\to \alpha)$
giving the decay rate for $b$ into that channel. We discuss the
relation between the spinor  representing $b$ and the channel
$\alpha$ in the
next section.

 $\cal M$ is a matrix
describing the evolution in the two-state space. For `cascade
mixing'   it is a product
of factors describing the evolution of the \sy as a \b, its
transition to a \k and finally its evolution as a \k. Since there 
no external disturbances (``decoherence") the entire process may be
simply regarded as a coherent evolution in the generalized
two-state
space. The time evolution factors are governed by the mass matrices
of the \b and
\k, and there is a `flip' amplitude A(1) at the time 1 giving the
transition from the
\b to \k. We will not be concerned with absolute
normalizations, so  only the
 structure of these  factors in the two-state space will
be of interest and multiplicative constants ignored. We finally
normalize the whole expression to some particular process.

  The  $\rho(a)$ and
$\rho(b)$ are density matrices
characterizing
the initial and final states.
 They arise
from the expression $\rho=vv^{\dagger}$, where $v$ is the `spinor'
of the two-state \sy decribing the initial or final state. The
definition of these states will be discussed in the next section.

Finally,  we note that by taking
the hermitian conjugate and using the permutation property of the
trace together with $\rho=\rho^\dagger$, one can
show that \eq{form} is always real. It is also  positive (or zero)
since it is the absolute value squared of a certain quantity,
namely $v^\dagger(b) {\cal M}\, v(a)$. Hermiticity of $\cal M$ is
not assumed.

\subsection{Definition of `Particle'}\label{ptle}
 It is important to recognize, as had been stressed in
ref\,\cite{alv}
that  a particular decay channel  can be used to
define some `\pl' in the two-state  \syn. Any decay
channel
$\alpha$
is described  by two complex numbers
$\alpha, \alpha'$,  giving the amplitude from say a \b or a \bb\,
into the given
channel. It is then possible to find two \b states, one {\it not}
going
into the
channel $ \alpha$ and one that {\it does} go. The one that does
{\it not}
decay can be constructed as  $\alpha'\ket{\bn}-\alpha\ket{\bbn}$,
since it will be seen
that the decay amplitude for this state $\sim
(\alpha\alpha'-\alpha'\alpha)=0$. On the other hand the orthogonal
state  $\alpha^*\ket{\bn}+\alpha'^*\ket{\bbn}$ {\it does} go into
the channel \footnote{The complex conjugates appearing here
explain why our basic expression \eq{form} has a somewhat different
ordering of the factors than the expression used in ref\,\cite{us}
(Eq 3). In that paper we used an approach based on amplitudes and
not state vectors, so that the corresponding state vector for a
given channel would have the complex conjugations. Here to avoid
possible misunderstandings we use the conventional language, where
a decay channel is characterized by a state vector, a certain \b or
\k
state, and so \eq{form} has the conventional form with the initial
state on the right and the
final state on the left.} $\alpha$. 
We have thus obtained two orthogonal states, one going and the
other not going into the given channel. Following ref\,\cite{alv}
one may call  these two
`\plsn' $B^\alpha$ and  $B^\alpha_\perp.$ In this way there is a
pair of \b states defined by every decay channel. The same of
course applies to  states in \k decay. 

In `measurement theory' language, if one thinks of the decay as a
`measurement' of the state of the `spinor', then that state which
does decay (like $B^{\alpha}$)  is the eigenstate for the
`measurement'.
Naturally, if there is some conserved quantum number such as CP,
various decay channels carrying this quantum number can in fact
define the same state of the two-state \syn.

Evidently, if  in \eq{form} one uses for $b$ that state 
which does decay into $\alpha$, then multiplying by the rate
constant
$\Gamma(b\to \alpha)$ gives the experimental rate. Thus
experimental
rates at different times for a given channel may be compared by
simply using  \eq{form} with the `eigenspinor' for the decay. To
compare different final channels in an absolute manner, knowledge
of the different `eigenspinors' and partial $\Gamma's$ is
necessary.

Finally, a signficant point about  $B^\alpha_\perp$  is that, for
a p-wave
pair as in $\Upsilon(4S)$ decay, if one side 
 is a $B^\alpha$,  then the other side is  necessarily a
$B^\alpha_\perp$.
This just follows from the linear \qm of a two-state \sy
with Bose-Einstein statistics and
do not involve any  symmetry properties such as CP. Indeed, without
further assumptions or information there is no
definite relation between the decays of $\ket{B_{\alpha}}$ and
those of $\ket{B_{\alpha\perp}}$, except as said, that 
$\ket{B_{\alpha\perp}}$ does not go into the channel $\alpha$.

\subsection{Density Matrices} 
 We shall use  density matrices in the following form:
\beql{rhoform}
 \rho(d)=\hf(1+{\bf d\cdot\sigma})\,
\eeql
where the $\bf \sigma$ are the three pauli matrices.
 One has $Tr\rho=1,$ reflecting the normalization of the state to
one. 
We use standard notation where $\sgz$ corresponds to the flavor
direction, 
$\sgz \ket{B^o}=+\ket{B^o}$, $\sgz \ket{\bar B^o}=-
\ket{\bar B^o}$, while $\sgy$ is the pure imaginary and anti-
symmetric matrix. Some definitions and relations are given in
the Appendix.

Furthermore we take $\bf d$ to be a  unit vector, ${\bf d}^2=1$.
This gives the properties

\beql{rhod}
 \rho^2(d)=\rho(d)~~~~~~~~~~~~~~~~~~~~~~~~~~~~\rho(d)\rho(-d)=0\,
\eeql

The first of these properties reflects the fact that we will always
have to do with `pure' states in the following \cite{groro}. 
The \dm for  the state $\ket{B_{\alpha\perp}}$ has the  
$\bf d$ opposite to that for
$\ket{B_{\alpha}}$.  This is
the meaning  of the second relation in Eq\,\ref{rhod}.

We now consider some definite states of interest and list the
properties
of the associated \dm in Table\,\ref{rtab}.
The first entries, for \b and \bb\, could be determined by a flavor
tag, as in leptonic decays of the type $B^o\to l^+...$. When the
\dm refers to an initial state determined by an `away
side' tag in $\Upsilon(4S)$ decay, then this initial state has the
opposite  $\bf d$ to the tag.

Next we can consider tags involving \jp\k and \jp\kb. In principle
this decay
amplitude has four basic possibilities according to whether \b or
\bb\, decays and whether the kaon is \k or \kb. However in the \sm
particle to antiparticle processes like $B^o\to J/\psi$\kb and
\bb$\to J/\psi$\k involve higher order weak transition and are
expected to be very small compared to the other two. We thus
neglect these, leaving the \b$\to J/\psi$\k and \bb$\to J/\psi$\kb
amplitudes.

In the tags involving \k's we shall neglect CP violation in
the \k \syn,
so for example  a $K_s$
refers to the \dtn of $\pi\pi$. Thus our discussion in these cases
can be inaccurate at the $10^{-3}$ level.

We now come to the next lines of Table\,\ref{rtab}, involving
\jp\ks and
\jp\kl. Without any particular assumptions these channels may be
taken as defining
two `\pls' $B^\alpha$ and $B^\beta $ and their orthogonal states,
as discussed in section\,\ref{ptle}.  

In principle the amplitudes for these channels could be found by
taking $\pm$ combinations of the \k,\kb\, amplitudes. However,
these
amplitudes are not completely known. Although the \k and \kb
amplitudes
refer to CPT conjugates, it is not permissable to use CPT for a
single channel in a many-channel situation. Hence it is
not possible to say more about the \jp\ks or \jp\kl channels
without further information or assumptions.

\subsection{Neglect of Direct CP Violation} \label{ngl}
Direct CP Violation is expected to be small for \b $\to$ \jp\k
and experimentally it is below the few percent level
\cite{bfac}. If we permit ourselves to neglect it in the
following and  further neglect all CP violation in the \k
\sy ($10^{-3}$ level), a great simplification ensuses. 
All CP violation arises through the mixing in the 
\b time evolution. Then the states defined by various tags 
may simply be given their naive CP assignments -- since no time
evolution is involved.

 Thus a \ks, identified by $\pi\pi$ decay,  is approximately the CP
even $K_1$. The tag
\jp\ks, involving the parity odd l=1,
identifies the parent \b as the \pl we may call the  CP odd $B_2$.

Then the $B^\alpha$ Table\,\ref{rtab} is 
simply a
$B_1$ and  $B^\alpha_\perp=B^\beta$ is the CP odd $B_2$. We show
these designations in the last
lines of the table. These identifications are certainly
approximate, but probably good
to the percent level. As we shall see below certain
simple relations in the simple one-time \b mixing problem follow
from this
assumption, and so larger
violations of these
relations can be taken as a suggestion of direct CP violation.

In using the naive CP assignments to make an  `away side' tag of
definite CP  it is of course not necessary that the tag be \jp\k,
but it must also  be one where direct CP violation is small.

\begin{table}
\begin{center}
\begin{tabular}{|l|l|l|l|}
\hline
Particle&Tag&${\bf d}=(d_1,d_2,d_3)$& Assumption\\
\hline
\hline
\b&$l^+...$&(0,0,\,1)&Standard\,model\\
\hline
\bb&$l^-...$&(0,0,-1)&Standard\,model\\
\hline
\b&\jp\k &(0,0,\,1)&Neglect  particle$\to$ antiparticle\\
\hline
\bb&\jp\kb &(0,0,-1)&Neglect particle$\to$ antiparticle\\
\hline
$B^\alpha$&\jp\ks &$(d_1,d_2,d_3)$&None\\
\hline
$B^\beta$&\jp\kl &$(d_1,d_2,d_3)$&None\\
\hline
$B_1$&\jp\kl &(1,0,0)&Neglect\;direct\,CP\,violation\\
\hline
$B_2$&\jp\ks &(-1,0,0)&Neglect\;direct\,CP\,violation\\
\hline
\end{tabular}

\end{center}
\caption{Values of $\bf d$ for some decay channels under different
assumptions. As explained in
the text, a `particle', that is, a certain linear combination in
the two-state \syn, may be defined by a decay channel. In
$\Upsilon(4S)\to$\b\bb\, when determining the initial state
by an
`away side' tag,  one has  $\bf -d$  for the `same side'. }
\label{rtab}
\end{table}

 \section{Simple \b Mixing }
We first consider the time evolution within the \b \sy only.
This will help in establishing the method and  showing the
relevant parameters.
 This is
of course a much studied subject \cite{bfac} , and we will mostly
reproduce known
results in the present language. 
  In the notation of \cite{us}, the
time evolution of the \b \sy is given by $S(1,0)=e^{-
iM_B\tau(1,0)}$, where $M_B$ is the complex, not necessarily
hermitian, mass matrix. For the purposes of this section
with only one time, we can call the time variable simply $\tau$ and
our basic expression \eq{form} becomes
\beql{forma}
 Tr[\rho(b)e^{-
iM_B\tau}\rho(a)(e^{-
iM_B\tau})^\dagger]\,
\eeql

 A great simplification,
as compared with the \k \syn, ensues here  in that the
non-hermitian part of
$M_B$, called $\hf \Gamma_B$, may be taken as proportional to the
identity matrix. That is, to a good approximation  \cite{acc} one
has  $\Delta \Gamma_B \approx 0$. With this
approximation
 $\Gamma$  factors out  of the
exponential, leaving a unitary matrix $U'$.

\beql{s}
S= e^{-\frac{\Gamma_B}{2} \tau} U'
~~~~~~~~~~~~~~~~~~~~~~~~~\Delta \Gamma_B\approx 0\,,
\eeql
and $U'$ gives simple unitary  
`rotations' in the \b \sy:
\beql{u}
U'=e^{-iM_B^H\tau}\,,
\eeql
where $M_B^H$ is the hermitian part of the mass matrix
$M_B^H=\hf(M_B+M_B^\dagger)$.

 The fact that the time evolution is  essentially  
given by a unitary matrix allows for  simplification of
the basic expression \eq{forma}. Expanding the two $\rho$ gives
four terms.
The `1' term gives simply $\hf$. The linear in ${\bf
d}_b\cdot\sigma$ terms
give zero using 
$U'U'^\dagger=1$ and $Tr \sigma=0$, leaving

\beal{passtr}
Tr[\rho(b)S\rho(a)S^\dagger]=
e^{-\Gamma_B\tau}\biggl(
\frac{1}{2}+\frac{1}{4}Tr\bigr[({\bf d}_b\cdot\sigma) U'({\bf
d}_a\cdot\sigma) U'^\dagger\bigl]\biggr)\\
\nonumber
~~~~~~~~~~~~~~~~~~~~~~~~~\Delta \Gamma_B \approx 0\,,
\eeal

With  the assumption of CPT invariance for the mass matrix the
diagonal 
elements of $M$ are equal, so that
\beql{m}
M_B^H= m_B^{av}I+ m_B=m_B^{av}+m_1\sgx +m_2\sgy\,
,~~~~~~~~~~~~~~~~~~~~~CPT\;
good
\eeql
with $ m_B^{av}$  the average mass of the \b.
The traceless part of the hermitian mass matrix
\beql{traceless}
m_B=m_1\sgx+m_2\sgy 
\eeql
 will play the most important role in the following.
 For the mass
splitting $\Delta m_B$
we have  
\beql{dm}
\hf \Delta m_B= \sqrt{m_1^2+m_2^2},
\eeql
and finally 
\beql{sa}
S= e^{-(im_B^{av}+\hf \Gamma_B)
\tau}U~~~~~~~~~~~U=e^{-i m_B\tau}=
e^{-i(m_1\sgx+m_2\sgy) \tau}
\eeql
 and \eq{passtr} becomes

\beal{passtra}
Tr[\rho(b)S\rho(a)S^\dagger]=
e^{-\Gamma_B\tau}\biggl(
\frac{1}{2}+\frac{1}{4}Tr\bigr[({\bf d}_b\cdot\sigma) U({\bf
d}_a\cdot\sigma) U^\dagger\bigl]\biggr)\\
\nonumber
~~~~~~~~~~~~~~~~~CPT\,good,~~~\Delta \Gamma_B \approx 0\,,
\eeal

The  evaluation of the traces in various expressions may be
simplified by using the absence
of the $\sgz$ term in $U$ and the 
anticommutation of the $\sigma$ to give \eq{antico} of the appendix
\beql{antiu}
\sgz U(\tau) = U(-\tau)\sgz \,.
\eeql
Also, note $U^\dagger(\tau)=U(-\tau)$, independently of CPT.

\section{CP and T Asymmetries in Simple Mixing }

 For the further  discussion one  needs the values of $m_1$
and $m_2$ individually. The mass splitting $\Delta m_B$  gives
$(m_1^2+m_2^2)$ via \eq{dm}
and  is known \cite{booklet} to be,
$\Delta m_B=0.51\times 10^{-12}s=3.3\times 10^{-10}MeV$. As
explained next, the 
measurements for CP or T asymmetries give
$m_2$, so then
both m's are
determined.

 We thus proceed to describe CP and T tests in simple \b
mixing in our
formalism, with the assumptions of  neglecting both direct CP
violation and a possible CPT violation. 
With the neglect of direct CP violation, both CP and T violation
effects will arise from the $m_2$ term in $S$. As should be
expected from the CPT theorem the violation of one symmetry will
imply the violation of the other, via $m_2\neq 0$.

\subsection{CP Asymmetry} \label{cpa1}
 Here one has compared\cite{belle},\cite{babar}, as a function of
time, \b
and \bb\, going
to a common, presumably
CP self-conjugate, final state,
namely \jp\ks. Identifying  \jp\ks with a decay of $B_2$ as
explained earlier, one defines the
asymmetry

\beql{asycp}
{\cal A}=\frac{Rate(\bar{B^o}\to B_2)-Rate({B^o}\to
B_2)}{Rate(\bar{B^o}\to B_2)+Rate(B^o\to B_2)}
\eeql
for different time intervals.  

 Using 
\eq{passtra} with the $\bf d$ from the
first, second, and last lines of  Table 1, one has
\beql{asycpa}
{\cal A}=\frac{\frac{2}{4} Tr[\sgx U \sgz U^\dagger] }{\hf+\hf}=\hf
Tr[\sgx U \sgz U^\dagger]\,.
\eeql

Using \eq{antico} to write
  $\hf
Tr[\sgx  U(\tau) \sgz U^\dagger(\tau)]=\hf Tr[-i\sgy U(-\tau) 
U^\dagger(\tau)]$=\\$\hf Tr[-i \sgy U(-2\tau)]$, one has
\beql{asf}
{\cal A}=-\hf Tr[i \sgy U(-2\tau)]
\eeql

We now use \eq{bexp}
\beql{cpres}
{\cal A}=-\hf Tr[i \sgy U(-2\tau)]=-\frac {m_2}{\hf\Delta m_B} sin
\bigr(\Delta m_B\, \tau\bigl)=-\frac {m_2}{\sqrt{m_1^2+m_2^2}}
sin \bigr(\Delta m_B \tau\bigl)\,.
\eeql
As was to be expected, with neglect of direct CP violation the
result is proportional to $m_2$. The formula is of course in
agreement with standard results \cite{bfac} with neglect of
direct CP violation. We note that this result, and in particular
the
fact that $m_2/\sqrt{m_1^2+m_2^2}$ is less than or equal to one,
 follows essentially from the \qm of mixing  and is
independent of any definite model of CP violation.

Thus the parameters $m_1$ and $m_2$ needed for our description are
given by $\Delta m_B$ and the coefficient of $ sin \bigr(\Delta m_B
\tau)$ in the asymmetry \eq{asycp}.
Experimentally, our  $\frac{m_2}{\sqrt{m_1^2+m_2^2}}$ is usually
referred to in the context of the CKM model as $sin 2 \beta$, and
has\cite{booklet}  the value  0.68, suggesting --in an interesting
coincidence--that $m_1$ and $m_2$ are about equal.

\subsection{CP Asymmetry Relations}
Because of the simple structure (within our approximations) of the
formulas  it is easy to establish
relations between  various other asymmetries  like \eq{asycp}. Let
us
write  ${\cal A}(\bbn,\bn ;B_2)$ for 
the asymmetry of \eq{asycp}.

 First we can
consider replacing the \ks in the \jp\ks final state with a \kl so
that we have a $B_1$ as the final state. According to the last line
of Table 1 this
means we should replace $\sgx$ in \eq{asycpa} with $-\sgx$. Thus
\beql{as1}
{\cal A}(\bbn,\bn;B_1)\approx -{\cal A}(\bbn,\bn;B_2)
\eeql

Next, we can consider a reverse type of procedure where starting
with  a CP eigenstate, $B_1$ or $B_2$, we go to
 two different but conjugate states, \b or 
\bb. The  initial $B_1$ or $B_2$ must of course be established by
an `away side' tag.
In an obvious notation
\beql{asycpx}
{\cal A}(B_2;\bbn,\bn)=\frac{Rate( B_2\to\bar{B^o})-Rate( B_2\to
B^o)}{Rate( B_2\to\bar{B^o})+Rate( B_2\to B^o)}
\eeql

 This amounts to exchanging $\sgx$ and $\sgz$ in \eq{asycpa}.
Looking at \eq{cpres} we see that $i\sgy$ will be replaced by $-
i\sgy$. Thus
\beql{as1a}
{\cal A}(B_2;\bbn,\bn)\approx -{\cal A}(\bbn,\bn;B_2)
\eeql
For $B_1$, we should, according to  Table 1, just change the sign:
\beql{as2}
{\cal A}(B_1;\bbn,\bn)\approx -{\cal A}(B_2;\bbn,\bn)\approx +{\cal
A}(\bbn,\bn;B_2)
\eeql
All these relations are expected to hold as a function of time. 
The approximations made are $\Delta \Gamma_B \approx 0$ and neglect
of direct CP violation in the \b \syn.  Neglect of
CP violation in the \k \sy is also implied, insofar that
identification of a \b state involves the assignment of a definite
CP to a \k. 

 Because of the approximations, particularly that of 
neglecting direct CP violation in the \b \syn, the equalities may
only be good to about the percent level. Alternatively, a breakdown
of the relations may be used to look for violations of the
assumptions. 

\section{T Asymmetry}
An interesting test showing manifest T violation in the \k \sy  was
carried out by
the LEAR group  in the 90's \cite{lear} and recently
analogous  tests have been discussed and presented\cite{texp} for
the \b \syn.
In these tests a certain time evolution `forwards' and `backwards'
is compared, and a difference in the two rates is a manifest
violation of T. One thus defines the asymmetry

\beql{ast}
{\cal A}(B^a\to B^b)=\frac{Rate(B^a\to B^b)-Rate({B^b}\to
B^a)}{Rate(B^a\to B^b)+Rate({B^b}\to B^a)}
\eeql
According to \eq{passtr} the numerator here is

\beal{pastr}
\frac{1}{4}\biggl(Tr\bigr[({\bf d}_a\cdot\sigma) U'({\bf
d}_b\cdot\sigma) U'^\dagger\bigl]-Tr\bigr[({\bf d}_b\cdot\sigma)
U'({\bf d}_a\cdot\sigma) U'^\dagger\bigl]\biggr)
\eeal
A notable consequence of this expression is that
if the $B^b$ state is the orthogonal state to the $B^a$ state, so
that ${\bf d}_b=-{\bf d}_a$, then $\cal A$ is zero
\beal{asz}
{\cal A}(B^a\to B^a_\perp)=0
\eeal
 For LEAR  the comparison was between \k $\to$ \kb\, and \kb $\to$
\k,  involving  in fact orthogonal states. Thus the nonzero result
found there is due to the significant $\Delta \Gamma \neq 0$ in the
\k \syn.
 In the \b case however, the analogous asymmetry ${\cal A}(\bn\to
\bbn)$ should be essentially zero since $\Delta \Gamma_B = 0$ holds
to high accuracy \cite{acc}.  These points are in agreement with
ref\,\cite{wolf} where it was pointed out that a nonzero $\Delta
\Gamma$ is needed for tests of this type.

\subsection{ T Test Formulas  }

 For tests of the type \eq{ast} it is thus necessary to chose
channels where $B^a, B^b$ are not orthogonal states, and discussion
\cite{wolf}
has centered around $B^a=\bn, B^b=B_2$, with $B_2$ identified via
the \jp$K_s$ tag. 
In this case \eq{ast} becomes
\beql{asta}
{\cal A}(B^o\to B_2)=\frac{Rate(B^o\to B_2)-Rate({B_2}\to
B^o)}{Rate(B^o\to B_2)+Rate({B_2}\to B^0)}\,,
\eeql
where with the identification of the \jp\ks tag with $B_2$ we 
neglect  direct CP violation in the \b \sy and CP violation in the
\k \syn.
Employing \eq{passtr} with  ${\bf d_a}=(0,0,1)$ and ${\bf
d_b}=(-1,0,0)$ 
\beal{pastrv}
{\cal A}(B^o\to B_2)=\frac{1}{4}\,
\frac{-Tr\bigr[\sgx U\sgz U^\dagger\bigl]\,\,+\,\,Tr\bigr[\sgz
U\sgx U^\dagger\bigl]}{ 1+Tr\bigr[\sgx U\sgz
U^\dagger\bigl]+Tr\bigr[\sgz
U\sgx U^\dagger\bigl]}
\,,
\eeal
 we see we have to do with the expressions
$Tr\bigr[\sgz U\sgx U^\dagger\bigl]$ and  $Tr\bigr[\sgx U\sgz
U^\dagger\bigl]$.  These 
quantities are equal and of opposite sign as follows by using
\eq{antico} and
$U(-\tau)=U^\dagger(\tau)$. Then
\beal{pastrx}
{\cal A}(B^o\to B_2)=
\frac{1}{2} Tr\bigr[\sgz U\sgx U^\dagger\bigl]=
\frac{1}{2}Tr\bigr[i\sgy U(-2\tau)\bigl]\,.
\eeal
But  this is the same as \eq{asf}, up to the sign. Therefore
\beql{cpeqt}
{\cal A}(\bn \to B_2)\approx -{\cal A}(\bbn,\bn;B_2),
\eeql
which was
evaluated in \eq{cpres}. It should be remarked, however, that the
CP test of  \eq{asf} and the T test here are not the same
quantities. The CP  asymmetry need not vanish at $t=0$ when direct
CP violation is not neglected \cite{bfac}, while here the vanishing
is an identity, following from
$Tr[\rho(a)\rho(b)]=Tr[\rho(b)\rho(a)]$.   

As with the CP asymmetries of \eq{as1} and \eq{as2},  different
variants of \\ ${\cal A}(B^o\to B_2)$ are simply related
\cite{bernabeu}. Replacing
\b by \bb\,leads to a minus sign due to $\bf d\to -d$, and
similarly for changing $B_1$ to $B_2$:

\beql{trel}
{\cal A}(\bbn\to B_2)\approx{\cal A}(\bn\to B_1)\approx -{\cal
A}(\bbn\to B_1)\approx -{\cal A}(B^o\to B_2)
\eeql
All of these are experimentally distinct possibilities, with
different combinations of `away side' and `same side' tags and,  
within the approximations, are all  given by \eq{cpres}.
That the T asymmetries are equal or opposite to the CP
asymmetries is
not very surprising since, given the assumptions,  both originate
from the
same $m_2$ term in the \b time  evolution.

 The approximations used for \eq{cpeqt} and \eq{trel} are, as
before,
$\Delta \Gamma_B \approx 0$, CPT, neglect of CP
violation in the kaon \syn, and neglect of direct CP violation in
the \b \syn. We thus should expect the relations to hold  at
least to the percent level.
 Violation of the various equalities would indicate a breakdown of
the assumptions, most likely that of neglecting direct CP violation
in the \b \syn, and violation above the percent level could be
suggestive of new physics.

\section{Cascade Mixing}

Having established the formalism and some parameters. We come
finally to ``cascade mixing" where we study the behavior 
in the two times $\tau(1,0)$ and $\tau(2,1)$. As explained 
earlier, in  $\Upsilon(4s)\to \bn\bbn$  time 0 and the starting \b
state can be determined by the `away side' tag. Time `1' is the
time of the `same side' decay  to \jp\k and time `2' is the time 
of the final \k decay. In 
\eq{form},  $\rho(a)$ is the \dm for the initial state of the \b, 
 $\rho(b)$ that for the final state of the \k and $\cal M$ is
\beql{dots}
  S(2,1) A(1) S(1,0)\, ,
\eeql
S(1,0) giving the propagation in time from 0 to 1 of the two-state
\sy, A(1) the transition within the \sy induced by  the decay  to
\jp\k, and  S(2,1) gives the propagation in time from 1 to 2.
S(1,0) was established above with the parameters as discussed in
section \ref{cpa1}

Concerning S(2,1) the neglect of CP violation for the \k (and of
course good CPT) allows us to write the mass matrix as
$M^{av}_K+m_K$, where $m_K$ is a matrix proportional to $\sgx$, so
that
\beql{s21}
S(2,1)=
e^{-i M^{av}_K \tau(2,1)}e^{-i m_K \tau(2,1)}\,,
\eeql
with the first term a scalar and the second a matrix operator.
 In contrast to the \b situation, $\Delta \Gamma$ here, although
approximately proportional to $\sgx$, is significantly different
from zero and must be retained. The matrix  $m_K$ thus represents
the complex \k
masses, with the hermitian part of $m_K$ representing $\hf$ the
mass splitting of \kl, \ks and the antihermitian part half of the
lifetime
difference $\Delta \Gamma_K= \hf(\Gamma_s-\Gamma_l)$ (see
Appendix). Because of
the retention of the matrix $\Delta
\Gamma_K\sim \sgx$, we now longer have the evolution in terms of a
unitary matrix as in \eq{u}, with $UU^\dagger=1$. 

Instead we have
\begin{multline}\label{ud}
{\cal M}{\cal M}^\dagger=e^{-\Gamma^{av}_K \tau(2,1)}e^{-\Gamma_B
\tau(1,0)}\times\\
 \bigl( e^{-i m_K \tau(2,1)}A(1)e^{-im_B \tau(1,0)}   e^{+im_B
\tau(1,0)} A^\dagger(1) e^{+i m^\dagger_K \tau(2,1)}\bigr)\\
 =e^{-\Gamma^{av}_K \tau(2,1)}e^{-\Gamma_B
\tau(1,0)}\times\bigl( e^{-i m_K \tau(2,1)} e^{+i
m^\dagger_K \tau(2,1)}\bigr)\\ =
e^{-\Gamma^{av}_K \tau(2,1)}e^{-\Gamma_B
\tau(1,0)} \times e^{-\Delta \Gamma_K \sgx \tau(2,1)} \,,
\end{multline}
where in the next-to-last step we used $ m_B=m^\dagger_B$. We also
took $A(1)A^\dagger(1)=1$ as will be used next in \eq{a1form}.

We now proceed to find the analog of the simple \eq{passtr}. The
lifetime prefactor now becomes $e^{-\Gamma_B \tau(1,0)}e^{-
\Gamma^{av}_K \tau(2,1)}$. So that \eq{form} now is
\beal{form2}
e^{- \Gamma^{av}_K \tau(2,1)} e^{-\Gamma_B \tau(1,0)}\times
~~~~~~~~~~~~~~~~~~~~~~~~~~~~~~~~~~~~~~~~~~~~~~~~~~~~ \\
\nonumber
Tr\bigl[\rho(b)  e^{-i m_K \tau(2,1)}A(1)e^{-im_B \tau(1,0)}
\rho(a)e^{+i m_B \tau(1,0)} A^\dagger(1) e^{+i m_K^\dagger
\tau(2,1)}  \bigr]
\eeal
with $m_B=m_1\sgx +m_2\sgy$.

 Turning now to A(1), we use the  neglect of \pl- anti\pl
transitions to set the amplitudes for $\bn \to$\jp \kb
and $\bbn \to$\jp \k to zero. This implies that A(1) is a diagonal
operator in our two-state flavor basis, leaving a linear
combination of I
and $\sgz$ as possibilities. However, if
we continue with the approximation of neglecting direct CP
violation for the \b and all CP violation for the \k, which permits
us to make naive CP identifications, then
  only $B^1\to K^2$ and $B^2\to K^1$
amplitudes are allowed. This then excludes I as a component of A(1)
and we may set
\beql{a1form}
A(1)\sim \sgz\,.
\eeql

\subsection{Evaluation of `Cascade' Terms}
There are in principle 16 different simple experimental observables
for \eq{form2}. For  $\rho(a)$ the intial \b could be tagged as a
\b, \bb, $B_1$, or $B_2$, while for  $\rho(b)$ the final \k can be
the analogous \k, \kb, $K_1$, or  $K_2$.
We proceed by considering the four terms resulting from the product
of the two
$\rho=\hf(1+\bf{d\cdot\sigma})$

\subsubsection {`1' term}\label{sub1}
The `1' term simply leads to the product evaluated in
\eq{ud}, hence this contribution is, taking the trace,
$e^{- \Gamma^{av}_K \tau(2,1)} e^{-\Gamma_B \tau(1,0)}\times$
\beql{1term}\frac{1}{4}
Tr \bigl[ e^{-\Delta \Gamma_K \sgx \tau(2,1)}  \bigr]= \frac{1}{4}
(e^{-\Delta \Gamma_K  \tau(2,1)}+e^{+\Delta \Gamma_K  \tau(2,1)})
=\hf cosh(\Delta \Gamma_K  \tau(2,1))
\eeql

\subsubsection {linear in \bf{d} terms}\label{ss2}

Unlike the discussion for \eq{passtr}, these linear terms do not
vanish since for the \k  one deals with a non-unitary
 evolution with a nonzero $\Delta\Gamma$. It will be seen that all
terms in this
section are proportional to $sinh \Delta \Gamma_K$ and so would
vanish
in the limit $\Delta \Gamma_K=0$

There are two terms, for ${\bf d}_a\cdot \sigma$ and for ${\bf
d}_b \cdot \sigma$. For \b, $B_1$ etc we then only have the two
cases,  up to a sign, 
${ d}_3 \sgz$  and ${ d}_1 \sgx $, for each $\bf d$.
\vskip2mm

{\bf Linear in $d_b$}:
\vskip3mm

${\bf d_b}$: With only the `1' term from $\rho(a)$, \eq{form2}
simplifies using  $A^2(1)=1$, so we are
left with $Tr[({\bf d}_b\cdot \sigma) e^{-\Delta \Gamma_K \sgx
\tau(2,1)} ]$. A $\sgz$ term gives zero so there
is only a contribution from a $\sgx$ term:

{\bf Case ${ d}_3 \sgz$ }: $Tr =0$

{\bf Case $d_1 \sgx$}: Using \eq{rlv}, we obtain
\beql{sinh}
-\hf sinh(\Delta \Gamma_K \tau(2,1))
\eeql
\vskip4mm
{\bf Linear in $d_a$}:
\vskip3mm
Now turning to the  terms proportional to
${\bf d_a}$, with only the `1' term from $\rho(b)$, we need to find

\begin{multline} \label{db}
\fth Tr\bigl[ e^{-i m_K \tau(2,1)}\sgz e^{-im_B \tau(1,0)}
({\bf d}_a\cdot \sigma) e^{+i m_B \tau(1,0)} \sgz
e^{+im_K^\dagger \tau(2,1)}  \bigr]=\\
\fth Tr\bigl[  e^{-im_B \tau(1,0)}
 ({\bf d}_a\cdot \sigma) e^{+i m_B \tau(1,0)}  e^{+
\Delta\Gamma_K\sgx \tau(2,1)}\bigr]\, , 
\end{multline}
where in the last line we have used \eq{antico} to pass through the
$\sgz$. At this point we
need \eq{rlv},
which  we insert
 in the last line of \eq{db}, the $cosh$ term vanishes
by the unitarity of the $m_B$ expression, leaving
\beql{db1}
\fth Tr\bigl[  e^{-im_B \tau(1,0)}
 ({\bf d}_a\cdot \sigma) e^{+i m_B \tau(1,0)} \sgx \bigr] {sinh(
\Delta\Gamma_K \tau(2,1))}
\eeql
to be evaluated.

There are now the two cases $({\bf d}_a\cdot \sigma)=\sgz$ or $
\sgx $, corresponding to a flavor or CP eigenstate for the initial 
\b.

{\bf Case $d_3\sgz$}: 
 Here \eq{antico} can be used again to pass through
the $\sgz$, giving for the trace $\fth Tr\bigl[  e^{-2im_B
\tau(1,0)}
  i\sgy \bigr] $.
This can be evaluated using \eq{bexp}, and we finally have
\beql{d3final}
\hf {sinh( \Delta\Gamma_K \tau(2,1))} \frac
{m_2}{\sqrt{m_1^2+m_2^2}}
sin \bigr(\Delta m_B\, \tau(1,0)\bigl)\, ,
\eeql
so that this term is proportional to the CP and T violating
parameter $m_2$

{\bf Case $d_1 \sgx$}: Here in \eq{db1} we have to do with the
expression
$\sgx e^{+i m_B \tau(1,0)} \sgx$. Using \eq{anticoa},  \eq{db1}
becomes
\begin{multline}\label{db2}
\fth Tr\bigl[  e^{-im_B \tau(1,0)}
  e^{+i {\tilde m_B} \tau(1,0)} \bigr] {sinh( \Delta\Gamma_K
\tau(2,1))}=\\
\hf  {sinh( \Delta\Gamma_K \tau(2,1))}\biggl( cos^2 \bigr(\hf
\Delta m_B\,
\tau(1,0)\bigl) +\biggl(\frac{m_1^2-m_2^2}{m_1^2+m_2^2}\biggr)
sin^2 \bigr(\hf \Delta m_B\, \tau(1,0)\bigr)\biggr) \, .
\end{multline}

 A somewhat simpler form results if we  add and subtract 1 so that
\eq{db2}
is
\beq\label{db3}
\hf {sinh( \Delta\Gamma_K \tau(2,1))}
\biggl(1+\biggl(\frac{-2m_2^2}{m_1^2+m_2^2}\biggr)  
sin^2 \bigr(\hf \Delta m_B \tau(1,0)\bigr)\biggr) \, ,
\eeq

In using the above results, it should be kept in mind that the sign
in front is relevant; thus ${\bf d}=(1,0,0)$  can correspond to a
$B_1$ while ${\bf d}=(-1,0,0)$  can correspond to a $B_2$ and so
on.

\subsubsection{$d_a d_b$ term}\label{ss3}
We now come to the last and most complicated term, where the trace
in \eq{form2} is
\beql{cpld}
\fth Tr\bigl[({\bf d}_b\cdot\sigma )  e^{-i m_K \tau(2,1)}\sgz
e^{-im_B \tau(1,0)}({\bf d}_a\cdot\sigma )
e^{+i m_B \tau(1,0)} \sgz e^{+i m_K^\dagger
\tau(2,1)}  \bigr]
\eeql
Since we consider flavor or CP tags, we have 4 possibilities here:
($\sgz,\sgz$), ($\sgz,\sgx$) ($\sgx,\sgz$) and ($\sgx,\sgx$). 

{\bf Term} ($\sgz,\sgz$)
 By repeated use of \eq{antico} \eq{cpld} can be reduced to
\begin{multline}\label{ss}
\fth Tr\bigl[  e^{+i m_K \tau(2,1)} e^{-im_B \tau(1,0)}
e^{-i m_B \tau(1,0)}  e^{+i m_K^\dagger
\tau(2,1)}  \bigr]=\\
\fth Tr\bigl[  e^{-i2 m_B \tau(1,0)}
  e^{+i(m_k+ m_K^\dagger)
\tau(2,1)}  \bigr]=\fth Tr\bigl[  e^{-i2 m_B \tau(1,0)}
  e^{+i\Delta m_K\sgx \tau(2,1)}  \bigr]\\
\end{multline}
Expanding $e^{+i\Delta m_K\sgx \tau(2,1)}=cos(\Delta m_K
\tau(2,1))+i\sgx sin(\Delta m_K \tau(2,1))$ and using \eq{cpres}
gives finally 
\begin{multline}\label{ss1}
\hf cos(\Delta m_B \tau(1,0))\, cos(\Delta m_K \tau(2,1))\\
+\hf \frac{m_1}{\sqrt{m_1^2+m_2^2}}
sin \bigr(\Delta m\, \tau(1,0)\bigl)sin(\Delta m_K \tau(2,1)\bigr)
\end{multline}
Through the $sin sin$ term this expression is sensitive to the
relative sign of $\Delta m_B,\Delta m_K$; with $m_2=0$ it would be
simply 
$\hf cos(\Delta m_B \tau(1,0) - \Delta m_K \tau(2,1))$, as was
found in
ref\,\cite{us} with CP conservation.

{\bf Term} ($\sgx,\sgx$)
Using \eq{antico}, \eq{cpld} is now reduced to
\beql{cpldz}
-\fth Tr\bigl[  e^{-i m_K \tau(2,1)}\sgy e^{-im_B
\tau(1,0)}\sgy
e^{-i m_B \tau(1,0)} e^{+i m_K^\dagger
\tau(2,1)}  \bigr]
\eeql
We now use \eq{anticoa}, $\sgy e^{-im_B
\tau(1,0)}\sgy=e^{+i\tilde{m_B}
\tau(1,0)}$, so that \eq{cpldz} is
\begin{multline} \label{cpldza}
-\fth Tr\bigl[  e^{-i m_K \tau(2,1)}
e^{+i\tilde{m_B}\tau(1,0)} e^{-im_B \tau(1,0)}
 e^{+i m_K^\dagger \tau(2,1)}  \bigr]=\\
-\fth Tr\bigl[e^{+i\tilde{m_B}\tau(1,0)}  e^{-im_B \tau(1,0)} 
 e^{- \Delta\Gamma_K\sgx \tau(2,1)}  \bigr]
\end{multline}
We  expand the $\Delta\Gamma $ term according to \eq{rlv}.
The coefficient of the $cosh$ term was evaluated in \eq{db2} so we
have
\begin{multline}\label{cosha}
-\hf {cosh( \Delta\Gamma_K \tau(2,1))}\biggl( cos^2 \bigr(\hf
\Delta m_B\,
\tau(1,0)\bigl) +\biggl(\frac{m_1^2-m_2^2}{m_1^2+m_2^2}\biggr)
sin^2 \bigr(\hf\Delta m_B\, \tau(1,0)\bigr)\biggr)=\\
-\hf {cosh( \Delta\Gamma_K \tau(2,1))}\biggl(
1+\biggl(\frac{-2m_2^2}{m_1^2+m_2^2}\biggr)
sin^2 \bigr(\hf\Delta m_B\, \tau(1,0)\bigr)\biggr) \, ,
\end{multline}

The coefficient of the $sinh$  involves $Tr[(m-{\tilde m})\sgx]$
and
$Tr[ m{\tilde m}\sgx] $, both of which are zero, so \eq{cosha} is
the only contribution to this term.

{\bf Term} ($\sgx,\sgz$)

Here we need 
\begin{multline}\label{s1s3}
\fth Tr\bigl[\sigma_1   e^{-i m_K \tau(2,1)}\sgz
e^{-im_B \tau(1,0)}\sigma_3
e^{+i m_B \tau(1,0)} \sgz e^{+i m_K^\dagger
\tau(2,1)}  \bigr]=\\
-\fth Tr\bigl[i\sgy
e^{-i2 m_B \tau(1,0)}  e^{-\Delta\Gamma_K\sgx
\tau(2,1)}  \bigr]\,.
\end{multline}
Expanding the two exponentials via \eq{bexp}  one finds the traces
of $\sgy$, $\sgy\sgx$, $\sgy\sgy\sgx$ or$\sgy\sgx\sgx$, all of
which are zero. The only nonzero factor is that of  $m_2\sgy\sgy$,
so this
term contributes
\beql{s1s2f}
-\hf \frac{m_2}{\sqrt{m_1^2+m_2^2}}
cosh\Delta\Gamma_K \tau(2,1))\,sin(\Delta m_b \tau(1,0))\,.
\eeql

{\bf Term} ($\sgz,\sgx$)

Here we need 
\begin{multline}\label{s1s3a}
\fth Tr\bigl[\sgz e^{-im_K\tau(2,1)} \sgz 
e^{-im_B \tau(1,0)}\sigma_1
e^{+i m_B \tau(1,0)} \sgz  e^{+im^\dagger_K\tau(2,1)} \bigr]=\\
\fth Tr\bigl[
e^{-i m_B \tau(1,0)} \sgx e^{+i m_B \tau(1,0)}\sgz  e^{+i(m_K+
m_K^\dagger)
\tau(2,1)} \bigr]=\\
-\fth Tr\bigl[
e^{-i { m_B} \tau(1,0)} i\sgy  e^{-i m_B \tau(1,0)}  e^{+i(m_K+
m_K^\dagger)
\tau(2,1)} \bigr]\,.
\end{multline}
We expand the $m_K$ exponential, the $sin$ term giving
\begin{multline}\label{s3s1}
-\fth sin(\Delta m_K\tau(2,1))Tr\bigl[i\sgx  e^{-i m_B
\tau(1,0)}i\sgy e^{-i m_B \tau(1,0)}
\bigr]=\\
\fth sin(\Delta m_K\tau(2,1))Tr\bigl[i\sgz  e^{+i{\tilde m_B}
\tau(1,0)} e^{-i m_B
\tau(1,0)}\bigr]=\\
\fth sin(\Delta m_K\tau(2,1))\frac{sin^2(\hf \Delta m_B \tau(1,0))}
{m_1^2+m_2^2} Tr\bigr[i\sgz  {\tilde m_B} m_B\bigl]=\\
-\frac{m_1m_2}{m_1^2+m_2^2}
  sin(\Delta m_K\tau(2,1))sin^2(\hf \Delta m_B \tau(1,0)) 
\,,
\end{multline}
while the $cos$ term is 
\begin{multline} \label{s3s1a}
\fth cos(\Delta m_K\tau(2,1))Tr\bigl[
e^{-i { m_B} \tau(1,0)}i\sgy  e^{-i m_B \tau(1,0)} \bigr]=\\
\fth cos(\Delta m_K\tau(2,1))Tr\bigl[i\sgy
e^{+i {\tilde m_B} \tau(1,0)}  e^{-i m_B \tau(1,0)} \bigr] =\\
\fth cos(\Delta m_K\tau(2,1))\frac {sin(\hf \Delta m_B
\tau(1,0))\,cos(\hf \Delta m_B \tau(1,0))}{\sqrt{m_1^2+m_2^2}}
Tr\bigl[\sgy (m_B-{\tilde
m_B} )  \bigr]=\\
\hf \frac{m_2}{\sqrt{m_1^2+m_2^2} }cos(\Delta m_K\tau(2,1))
\,sin(\Delta m_B \tau(1,0))\,.
\end{multline}

\begin{table}
\begin{tabular}{|l|l|l|}
\hline
Name&Function&reference\\
\hline 
\hline
A&$cosh \Delta \Gamma_K\tau(2,1)$&\eq{1term};\eq{cosha} \\
\hline
B&$sinh \Delta \Gamma_K\tau(2,1)$&
 \eq{sinh}; \eq{db3} \\
\hline
C&$sinh \Delta\Gamma_K\tau(2,1)\, sin \Delta
m_B\tau(1,0) $ &
 \eq{d3final} \\
\hline
D&$sinh \Delta \Gamma_K\tau(2,1)\, sin^2 \hf \Delta             
 m_B\tau(1,0)$&
 \eq{db3} \\
\hline
E&$cos \Delta m_K\tau(2,1) \,
cos \Delta m_B\tau(1,0)$&
 \eq{ss1} \\
\hline
F&$sin \Delta               
m_K\tau(2,1) \, sin \Delta m_B\tau(1,0) $&
 \eq{ss1} \\
\hline
G&$cosh \Delta \Gamma_K\tau(2,1)\,
 sin^2 \hf \Delta m_B\tau(1,0))$&
 \eq{cosha} \\
\hline
H&$cosh \Delta \Gamma_K\tau(2,1)\,
 sin \Delta m_B\tau(1,0)$ &
 \eq{s1s2f} \\
\hline
I&$cos\Delta m_K\tau(2,1)
\, sin \Delta m_B\tau(1,0)$&
\eq{s3s1a} \\
\hline
J& $sin \Delta m_K\tau(2,1)
\, sin^2 \hf \Delta m_B\tau(1,0)$&
 \eq{s3s1} \\
\hline
\end{tabular}
\caption{Functions employed with their designations and reference
in the text.}
\label{ftns}
\end{table}
\subsection{Tabulation}
 We now have completed the evaluation of the terms that will appear
in the cascade mixing of all 16 combinations of initial and
final tags. We present these in tabular form. Table \ref{ftns} 
lists the  functions that occur.  They are normalized
so that at $\tau(2,1)=\tau(1,0)=0$, they are equal to either zero
or one.
A given process, beginning with a \b and ending with a \k, called
'a' and 'b' respectively,
will be given by sums of these functions with
different coefficients. The
coefficients for each process are given in Table\,\ref{ftab}.

If desired, the factors $ sin^2 \hf \Delta m_B\tau(1,0))$  can be
expanded using half angle identities to give full angle
expressions; thus combinations of the
functions I and J can be made to exhibit expressions of the type  
 $sin(\Delta m_B \tau(1,0)+\Delta m_K\tau(2,1))$, where the
relative sign of the $\Delta m$ enters, as was noted in
ref\,\cite{us}. Similarly E and F when combined in the limit
$m_2=0$ give $cos(\Delta m_B \tau(1,0)-\Delta m_K\tau(2,1))$.   One
may also multiply the $sinh  \Delta
\Gamma_K\tau(2,1)$ or $cosh  \Delta
\Gamma_K\tau(2,1)$ factors by the prefactor $e^{
\Gamma^{av}_K\tau(2,1)}$  to exhibit the contributions 
 of the two \k lifetime eigenstates.
While $\Delta m_K$ occurs in the functions
of Table \ref{ftns}, the coefficients in Table \ref{ftab},  depend
only on the parameters of the \b
mass matrix. This is because with the assumption of  neglect of CP
violation in the \k \syn, there is essentially only  one paramter,
the coefficient of $\sgx$, and one has simply $e^{-i\Delta
m_K\sigma_1\tau}=cos(\Delta m_K\tau)-i\sigma_1sin(\Delta m_K\tau)$.
\begin{landscape}
\begin{table}
\renewcommand{\arraystretch}{1.2}
\begin{tabular}{|l|l|l|l|l|l|l|l|l|l|l|l|l|l|l|}
\hline
a&b&${\bf d}_b$&${\bf d}_a$&A&B&C&D&E&F&G&H&I&J\\
\hline
\hline
\b&\k&(0,0,\,1)&(0,0,\,1)&$\frac{1}{2}$&0&
$\hf \frac{m_2}{\sqrt{m_1^2+m_2^2}}$&0&$\frac{1}{2}$&$\hf\frac{m_1}
{\sqrt{m_1^2+m_2^2}}$&0&0&0&0\\
\hline
\bb&\k&(0,0,\,1)&(0,0,-1)&$\frac{1}{2}$&0&
$-\hf \frac{m_2}{\sqrt{m_1^2+m_2^2}}$&0&$-\hf$&$-\hf
\frac{m_1}{\sqrt{m_1^2+m_2^2}}$&0&0&0&0\\
\hline
$B_1$&\k&(0,0,\,1)&(1,0,0)&$\frac{1}{2}$&$\hf$&0&$\frac{-
m^2_2}{m^2_1+m^2_2}$&0&0&0&0&$\hf\frac{m_2}{\sqrt{m_1^2+m_2^2}
}$&$-\frac{m_1m_2}{m_1^2+m_2^2}$\\
\hline
$B_2$&\k&(0,0,\,1)&(-1,0,0)&$\frac{1}{2}$&$-
\hf$&0&$\frac{m^2_2}{m^2_1+m^2_2}$&0&0&0&0&$-
\hf\frac{m_2}{\sqrt{m_1^2+m_2^2}}$&$
\frac{m_1m_2}{m_1^2+m_2^2}$\\
\hline
\b&\kb&(0,0,-1)&(0,0,\,1)&$\frac{1}{2}$&0&$
\hf \frac{m_2}{\sqrt{m_1^2+m_2^2}}$&0&$-\hf$&$-\hf\frac{m_1}
{\sqrt{m_1^2+m_2^2}}$ &0&0&0&0\\
\hline
\bb&\kb&(0,0,-1)&(0,0,-1)&$\frac{1}{2}$&0&$-
\hf \frac{m_2}{\sqrt{m_1^2+m_2^2}}$&0&$\hf$&$\hf\frac{m_1}
{\sqrt{m_1^2+m_2^2}}$ &0&0&0&0\\
\hline
$B_1$&\kb&(0,0,-1)&(1,0,0)&$\frac{1}{2}$&$\hf$&0&$\frac{-
m^2_2}{m^2_1+m^2_2}$&0&0&0&0&$-
\hf\frac{m_2}{\sqrt{m_1^2+m_2^2}}$&$
\frac{m_1m_2}{m_1^2+m_2^2}$\\
\hline
$B_2$&\kb&(0,0,-1)&(-1,0,0)&$\frac{1}{2}$&$-
\hf$&0&$\frac{m^2_2}{m^2_1+m^2_2}$&0&0&0&0&$\hf\frac{m_2}{\sqrt
{m_1^2+m_2^2}}$&$-\frac{m_1m_2}{m_1^2+m_2^2}$\\
\hline
\b&$K_1$&(1,0,0)&(0,0,\,1)&$\frac{1}{2}$&$-\hf$&$
\hf
\frac{m_2}{\sqrt{m_1^2+m_2^2}}$&0&0&0&0&$\hf\frac{m_2}{\sqrt{m_1^
2+m_2^2}}$&0&0\\
\hline
\bb&$K_1$&(1,0,0)&(0,0,-1)&$\frac{1}{2}$&$-\hf$&$-
\hf \frac{m_2}{\sqrt{m_1^2+m_2^2}}$&0&0&0&0&$-
\hf\frac{m_2}{\sqrt{m_1^2+m_2^2}}$&0&0\\
\hline
$B_1$&$K_1$&(1,0,0)&(1,0,0)&$\frac{1}{2}-\hf$&$-\hf+\hf$&0&$\frac{-
m^2_2}{m^2_1+m^2_2}$&0&0&$\frac{m^2_2}{m^2_1+m^2_2}$&0&0&0\\
\hline
$B_2$&$K_1$&(1,0,0)&(-1,0,0)&$\frac{1}{2}+\hf$&$-\hf -
\hf$&0&$\frac{m^2_2}{m^2_1+m^2_2}$&0&0&$\frac{-m^2_2}{m^2_1+m^2_
2}$&0&0&0\\
\hline
\b&$K_2$&(-1,0,0)&(0,0,\,1)&$\frac{1}{2}$&$\hf$&$
\hf \frac{m_2}{\sqrt{m_1^2+m_2^2}}$&0&0&0&0&$
-\hf\frac{m_2}{\sqrt{m_1^2+m_2^2}}$&0&0\\
\hline
\bb&$K_2$&(-1,0,0)&(0,0,-1)&$\frac{1}{2}$&$\hf$&$-
\hf
\frac{m_2}{\sqrt{m_1^2+m_2^2}}$&0&0&0&0&$\hf\frac{m_2}{\sqrt{m_1^
2+m_2^2}}$&0&0\\
\hline
$B_1$&$K_2$&(-1,0,0)&(1,0,0)&$\frac{1}{2}+\hf$&$\hf+\hf$&0&$\frac{-
m^2_2}{m^2_1+m^2_2}$&0&0&$\frac{-m^2_2}{m^2_1+m^2_2}$&0&0&0\\
\hline
$B_2$&$K_2$&(-1,0,0)&(-1,0,0)&$\frac{1}{2}-\hf$&$\hf-
\hf$&0&$\frac{m^2_2}{m^2_1+m^2_2}$&0&0&$\frac{m^2_2}
{m^2_1+m^2_2}$&0&0&0\\
\hline
\end{tabular}
\caption{Coefficients of functions arising in \b
$\to$(\jp$)\to$\k, for 16 possible combinations of initial \b and
final \k.   
To find the relative rates for a given process as function of
$\tau(1,0)$ and $\tau(2,1)$,
use the functions
given in Table \ref{ftns}, summed with the coefficents given here;
and
multiply by the prefactor $e^{-\Gamma_B\tau(1,0)}e^{-
\Gamma_K^{av}\tau(2,1)}$.
The normalization is such that the value
for \b $\to$\kb at $\tau(1,0)=\tau(2,1)=0$ is 1. 
Values of the $\bf d$ are given as $(d_1,d_2,d_3)$. Where a
coefficient
arises from two different equations in the text, we exhibit the
contributions individually. The \b mass splitting is
$2\sqrt{m_1^2+m_2^2}$
and $m_2$ gives the CP and T violation in the \b mass matrix.}
\label{ftab}
\end{table}
\end{landscape}

\section{Conclusions}
We have explained how using an `away side tag' can make it possbile
to observe `cascade mixing' in $\Upsilon(4S)\to \bn \bbn$ without
precision knowledge on the location of the primary vertex.

Confirmation of our predictions, as summarized in
Table\,\ref{ftab}, would verify the validity of this pretty
extension of the physics of particle mixing and also provide an
additional approach to the parameters of the \b mass matrix,
particularly the CP violating parameter, here called $m_2$.

Since, as explained in section \ref{cpa1}, it appears that $m_1$
and $m_2$ are of about the same size, the coefficients in Table
\ref{ftab} should  have
substantial values.

The role of the first flight time as a `variable
regenerator' depending on $\tau(1,0)$, can be nicely exhibited,
data permitting, by exhibiting the different oscillation patterns
arising for the \k according to the value of $\tau(1,0)$.
By setting $\Delta m_B\tau(1,0)$
in the vicinity of 0, $\pi$, $2\pi$...for example, various of the
functions in Table\,\ref{ftns} can be made to disappear or to
reverse sign, giving distinctly different patterns in $\tau(2,1)$.

Those processes involving the function F can give information about
the relative sign of $\Delta m_B,\Delta m_K$ (ref\,\cite{az}),
while the coefficients of function J involve the relative sign of 
$m_1,m_2$.

Our approximations include $\Delta \Gamma_B\approx 0$, good CPT,
neglect of CP violation in the \k \syn, neglect of \pl-anti\pl
processes, and
neglect of direct CP violation. The latter is probably the most
significant, and enters in two ways. One is that it allows the
use of specific decay channels to identify states of the \k or \b
\sys with their naive CP values, as discussed in section \ref{ngl}.
Secondly it allows for the simple form of the transition
amplitude A(1) in \eq{a1form}. Because of these assumptions, the
predictions can only be taken to  be
good to the percent level.

Many extensions and generalizations using the method can be
envisioned. These could include 
replacing the \jp which induces
the  conversion from \b to \k, with another \pl or \pl state, for
example one where the CP is not `flipped', or one where there is
direct CP violation, in which case the  transition operator
 \eq{a1form} would contain a mixture of terms.
One may consider introducing the $D^0$ \cite{branco}( most amusing 
would
be a `three-time cascade' \b $\to D^o\to$\k which could be treated
by the methods here) and undoubtedly many other possibilities.

\section{Acknowledgements}
I would like to thanks B. Kayser for many discussions on this topic
over the years, and C. Kiesling for much help in understanding the
current state of \b physics and the experimental possibilities.

\newpage

\section{Appendix}
 We list some of the definitions and $\sigma$ relations 
used in the text.
\subsection{Mass Matrices}
The mass matrices used, after removal of the average mass and
average $\Gamma$ are
\beql{am}
m_B=m_1\sgx +m_2\sgy~~~~~~~~~~~~~~~~~~~~m_K=\hf (\Delta m_K -
i\Delta \Gamma_K)\sgx \,,
\eeql
where the terms proportional to the I matrix, namely $-
i\hf\Gamma_B$, $-i\hf\Gamma^{av}_K$, as well as the average real
masses have been taken out. These definitions for $m_B$ and $m_K$
represent our approximations of neglecting $\Delta \Gamma_B$
and neglecting CP violation in the \k \syn. In terms of the
hermitian
and antihermitian parts for the \k

\begin{multline}\label{agam}
(m_K+m^\dagger_K)=\Delta m_K \sgx~~~~~~~~~~~~~~~~~~~-i(m_K-
m^\dagger_K)=-\Delta \Gamma_K \sgx\\
\end{multline}
In terms of the definite lifetimes
\begin{multline}\label{gamph}
\Gamma_K^{av}= \hf(\Gamma_s+ \Gamma_l)~~~~~~~~~~~~~~~~~~~~~~~\Delta
\Gamma_K= \hf(\Gamma_s- \Gamma_l)
\end{multline}

For the mass splitting $\Delta m_B$ of the \b
\beql{bspl}
\hf \Delta m_B= \sqrt{m_1^2+m_2^2}\,,
\eeql
while for the \k the mass splitting is simply $\Delta m_K$.

\subsection{\bf Pauli  matrices}

We use the definition of the pauli matrices where
\beql{pauli}
\sgx\sgy=i\sgz~~~~~~~~~~~~~~~~\sgx\sgy=-
\sgy\sgx~~~~~~~~~~~~~~~\sgx^2=I
\eeql
and cyclic permutations.

 We frequently use the identity
\beql{exp}
e^{-i{\bf b\cdot\sigma}}= cos\, b -i ({\bf 
b\cdot\sigma})\frac{sin\,b}{b}
\eeql 
which for the \b time development implies
\begin{multline}\label{bexp}
e^{-i m_B\tau}= cos(\hf\Delta m_B \tau ) -i\frac{ (m_1\sgx
+m_2\sgy)}{\hf\Delta m_B} sin(\hf\Delta m_B \tau )\\
e^{-i2 m_B\tau}= cos(\Delta m_B \tau ) -i\frac{ (m_1\sgx
+m_2\sgy)}{\hf\Delta m_B} sin(\Delta m_B \tau )
\end{multline}

We use the real version of \eq{exp} in the \k evolution as
\beql{rlv}
e^{-\Delta\Gamma_K \tau\sgx}=cosh(\Delta\Gamma_K \tau)- \sgx {sinh(
\Delta\Gamma_K \tau)}
\eeql

From the absence of $\sgz$ (CPT assumption) in the \b and \k time
evolutions one has, by anticommuting the $\sigma$. the useful
identities
\begin{multline}\label{antico}
\sgz e^{-im_B \tau}=e^{+im_B \tau}\sgz~~~~~~~~~~~~~~~~
\sgz e^{-im_K \tau}=e^{+im_K \tau}\sgz
\end{multline}
Analogous relations are
\begin{multline}\label{anticoa}
 \sgx e^{-im_B \tau}= \sgx e^{-i(m_1\sgx
+m_2\sgy) \tau}= e^{-i(m_1\sgx-m_2\sgy) \tau}\sgx=e^{-i\tilde{m_B}
\tau}\sgx\\
\sgy e^{-im_B \tau}=\sgy e^{-i(m_1\sgx
+m_2\sgy) \tau}= e^{+i(m_1\sgx-m_2\sgy) \tau}\sgy=e^{+i\tilde{m_B}
\tau}\sgy
\, ,
\end{multline}
where $\tilde{m_B}$ is the transpose of  ${m_B}$.

\end{document}